\documentclass[useAMS,usenatbib,usegraphicx]{mn2e}

\usepackage{indentfirst}
\usepackage{amssymb}

\newcommand{\D}[2]{\frac{\partial #2}{\partial #1}}
\newcommand\bb[1] {\mbox{\boldmath{$#1$}}}
\newcommand\del{\bb{\nabla}} 
\newcommand\bcdot{\bb{\cdot}}
\newcommand\btimes{\bb{\times}} 
\newcommand\real{{\rm Re}} 

\title[MRI-driven Reynolds and Maxwell stresses in accretion discs]
{The signature of the magnetorotational instability in the Reynolds
  and Maxwell stress tensors in accretion discs} 

\author[M.E. Pessah, C.K. Chan, and D. Psaltis] {Martin
  E. Pessah$^{1,2}$\thanks{E-mail:mpessah@as.arizona.edu (MEP)},
  Chi-kwan Chan$^{2}$, and Dimitrios Psaltis$^{2,1}$\vspace{6pt}\\
  $^{1}$Astronomy Department, 933 N. Cherry Ave., Tucson, AZ, 85721, USA\\
  $^{2}$Physics Department, 1118 E. $4^{th}$ St., Tucson, AZ, 85721,  USA}

\begin{document}
\date{Accepted 2006 July 19. Received 2006 July 18; in original form 2006 March 09.}

\pagerange{\pageref{firstpage}--\pageref{lastpage}} \pubyear{2006}

\maketitle

\label{firstpage}

\begin{abstract}
The magnetorotational instability is thought to be responsible for the
generation of magnetohydrodynamic turbulence that leads to enhanced
outward angular momentum transport in accretion discs.  Here, we
present the first formal analytical proof showing that, during the
exponential growth of the instability, the mean (averaged over the
disc scale-height) Reynolds stress is always positive, the mean
Maxwell stress is always negative, and hence the mean total stress is
positive and leads to a net outward flux of angular momentum.  More
importantly, we show that the ratio of the Maxwell to the Reynolds
stresses during the late times of the exponential growth of the instability
is determined only by the local shear and does not depend on the
initial spectrum of perturbations or the strength of the seed
magnetic.  Even though we derived these properties of the stress
tensors for the exponential growth of the instability in
incompressible flows, numerical simulations of shearing boxes show
that this characteristic is qualitatively preserved under more general
conditions, even during the saturated turbulent state generated by the
instability.  
\end{abstract}

\begin{keywords}
black hole physics -- accretion, accretion discs -- MHD -- instability
-- turbulence.
\end{keywords}


\section{Introduction}
\label{sec:intro}

Magnetohydrodynamic (MHD) turbulence has long been considered
responsible for angular momentum transport in accretion discs
surrounding astrophysical objects \citep{SS73}.  Strong support for
the importance of magnetic fields in accretion discs followed the
realization by \citet{BH91} that laminar flows with radially
decreasing angular velocity profiles, that are hydrodynamically
stable, turn unstable when threaded by a weak magnetic field. Since
the discovery of this magnetorotational instability (MRI), a variety
of local \citep[e.g.,][]{HGB95, HGB96, Brandenburg95, Brandenburg01,
  Sanoetal04} and global \citep[e.g.,][]{H00, H01, SP01} numerical
simulations have shown that its non-linear evolution gives rise to a
turbulent state characterized by enhanced Reynolds and Maxwell
stresses, which in turn lead to outward angular momentum transport.

The relevance of the Reynolds and Maxwell stresses in determining the
dynamics of a magnetized accretion disc is best appreciated by
examining the equation for the dynamical evolution of the mean
specific angular momentum of a fluid element \cite[see, e.g.,][]{BH98,
  BP99}.  Defining this quantity for a fluid element with mean density
$\bar{\rho}$ as $\bar l \equiv r \bar \rho {\bar v_\phi}$ we can write
in cylindrical coordinates
\begin{equation}
\label{eq:angular_momentum_mean}
\partial_t \bar l + \del \bcdot (\bar l \bar{\bb{v}}) = \frac{1}{4\pi}
\del \bcdot r (\bar B_{\phi} \bar{\bb{B}} ) + \frac{1}{4\pi} 
\del \bcdot r \overline{\delta\!B_\phi \delta\!\bb{B}} - 
\del \bcdot r \overline{\rho \delta\!v_\phi \delta\!\bb{v}} \,, \  
\end{equation}
where $\bar{\bb{v}}, \bar{\bb{B}}$, $\delta \bb{v}$, and $\delta
\bb{B}$ stand for the means and fluctuations of the velocity and
magnetic fields, respectively, and the bars denote suitable averages.
It is clear that the presence of mean magnetic fields or of
non-vanishing correlations between the fluctuations in the magnetic or
velocity field can potentially allow for the specific angular momentum
of a fluid element to change.  In the absence of strong large-scale
magnetic fields, the last two terms on the right hand side will
dominate and we can simplify equation (\ref{eq:angular_momentum_mean})
as
\begin{equation}
\label{eq:angular_momentum_mean_weak}
\partial_t \bar l + \del \bcdot (\bar l \bar{\bb{v}}) =  - \del \bcdot
(r \bar{\bb{\mathcal F}}) \,,
\end{equation}
where the vector $\bar{\bb{\mathcal F}}$ characterizes the flux of
angular momentum. Its components are related to the Reynolds
and Maxwell stresses,
$\bar R_{ij} = \overline{\rho \, \delta\!v_i \, \delta\!v_j}$ and
$\bar M_{ij} = \overline{\delta\!B_i \, \delta\!B_j}/4\pi$ via
\begin{equation}
\bar{\mathcal F}_i \equiv  \bar R_{i\phi} - \bar M_{i\phi} = \bar T_{i\phi}
\,.
\end{equation}
It is straightforward then to see that in order for matter in the disc
to accrete, i.e., to loose angular momentum, the sign of the mean
total stress, $\bar T_{r\phi}$, must be positive.  Note that in the
MRI-literature the Maxwell stress is defined as the negative of the
correlations between magnetic field fluctuations. We have chosen
instead to use a definition in which the sum of the diagonal terms
has the same sign as the magnetic energy density.

In a differentially rotating MHD turbulent flow, the sign of the
stress component $\bar T_{r\phi}$ will depend on the mechanism driving
the turbulence (presumably the MRI), the mechanism mediating the
energy cascade between different scales, and the dissipative processes
that lead to saturation. Nevertheless, mechanical analogies of the MRI
\citep{BH92, BH98, KFM98, BC97, B03} as well as analyses involving the
excitation of single-wavenumber modes \citep[][see also \S 4]{BH92,
BH02, NQIA02} suggest that angular momentum in transported outwards
even during the linear phase of the instability.

In this paper, we derive analytic expressions that relate the dynamics
of the MRI-driven fluctuations in Fourier space with the mean values,
averaged over the disc scale-height, of the different stresses in
physical space.  This allows us to provide the first formal analytical
proof showing that the MRI leads to a positive mean total stress $\bar
T_{r\phi}$ which, in turn, leads to a net outward transport of angular
momentum.  Within this formalism, we demonstrate that the Reynolds
stress is always positive, the Maxwell stress is always negative, and
that the absolute value of the Maxwell stress is larger than the
Reynolds stress as long as the flow in the absence of magnetic fields
is Rayleigh-stable.  Moreover, we uncover a robust relationship
between the Maxwell and Reynolds stresses as well as between the
magnetic and kinetic energy of the MHD fluctuations during the late
times of the linear phase of the instability. Specifically, we show
that the ratio of the (absolute value of the) Maxwell and Reynolds
stresses is equal to the ratio of magnetic and kinetic energy
densities, and that both ratios depend only on the value of the local
shear characterizing the flow.

The rest of the paper is organized as follows.  In \S 2 we state our
assumptions.  In \S 3 we present the complete solution to the
MRI-eigenvector problem in Fourier space.  We pay particular attention
to the characterization of the complex nature of the MRI-eigenvectors.
In \S 4, we present the formalism to derive analytic expressions for
the mean physical Reynolds and Maxwell stresses in terms of the
fluctuations in spectral space.  We show that the MRI leads to a net
outward transport of angular momentum. We also study there the
different properties of the MRI-driven stresses during the late times
of their exponential growth. In \S 5 we compare these stress
properties with similar properties found in previous numerical
simulations that addressed the non-linear turbulent regime using
shearing boxes.  We also present there our conclusions.

\section{Assumptions}
\label{sec:assumptions}

This paper is concerned with the signature of the axisymmetric MRI in
the mean values (averaged over the disc scale-height) of the Reynolds
and Maxwell stress tensors.  In particular, we consider a cylindrical,
incompressible background, characterized by an angular velocity
profile $\bb{\Omega}=\Omega(r)\check{\bb{z}}$, threaded by a weak
vertical magnetic field $\bb{\bar{B}} = \bar{B}_z \check{\bb{z}}$.  In
order to address this issue, we work in the shearing sheet
approximation, which has proven useful to understand the physics of
disc phenomena when the scales involved are smaller than the disc
radius.

The shearing sheet approximation consist of a first order expansion in
the variable $r-r_0$ of all the quantities characterizing the flow at
the fiducial radius $r_0$.  The goal of this expansion is to retain
the most important terms governing the dynamics of the MHD fluid in a
locally-Cartesian coordinate system co-orbiting and corrotating with
the background flow with local (Eulerian) velocity $\bb{v} =
r_0\,\Omega_0 \check{\bb{\phi}}$. (For a more detailed discussion
regarding the shearing sheet approximation, see \citealt{GX94} and
references therein.)

The equations for an  incompressible MHD flow in
the shearing sheet limit are given by
\begin{eqnarray}
\label{eq:euler}
\D{t}{\bb{v}} + \left(\bb{v}\bcdot\del\right)\bb{v} & = &
- 2 \bb{\Omega}_0 \btimes \bb{v} \, + 
\, q \Omega^2_0\del(r-r_0)^2 \nonumber \\
& & - \frac{1}{\rho}\del\left(P + \frac{\bb{B}^2}{8\pi}\right)  +
\frac{(\bb{B}\bcdot\del)\bb{B}}{4\pi\rho}  \\
\label{eq:induction}
\D{t}{\bb{B}} + \left( \bb{v} \bcdot \del \right)\bb{B} 
& = & \left(\bb{B} \bcdot \del \right) \bb{v} \, 
\end{eqnarray}
where $P$ is the pressure, $\rho$ is the (constant) density, the
factor $q\equiv-d\ln\Omega/d\ln r|_{r_0}$ parametrizes the magnitude
of the local shear, and we have defined the (locally-Cartesian)
differential operator
\begin{eqnarray}
\del & \equiv &  
\check{\bb{r}} \, \frac{\partial}{\partial r}  + 
\frac{\check{\bb{\phi}}}{r_0}\,\frac{\partial}{\partial \phi} +
\check{\bb{z}} \, \frac{\partial}{\partial z} \,,
\end{eqnarray}
where $\check{\bb{r}}$, $\check{\bb{\phi}}$, and $\check{\bb{z}}$ are,
coordinate-independent, orthonormal vectors corrotating
with the background flow at $r_0$.

In what follows, we focus our attention on fluctuations that depend
only on the vertical coordinate.  These types of fluctuations are
known to have the fastest growth rates \citep{BH92, BH98} and will,
therefore, constitute the most important contributions to the
corresponding Reynolds and Maxwell stresses during the exponential
growth of the instability.  The equations governing the dynamics of
these fluctuations can be obtained by noting that the velocity and
magnetic fields given by
\begin{eqnarray}
\label{eq:mean_plus_fluctuations_v}
\bb{v} &=& 
\delta v_r(z) \check{\bb{r}} + [- q \Omega_0 (r-r_0)+ \delta v_\phi(z)] 
\check{\bb{\phi}} + \delta v_z(z) \check{\bb{z}}  \,, \\
\label{eq:mean_plus_fluctuations_b}
\bb{B} &=& \delta B_r(z) \check{\bb{r}}  + 
\delta B_\phi(z) \check{\bb{\phi}} + 
[\bar B_z + \delta B_z(z)] \check{\bb{z}} \,,
\end{eqnarray}
constitute a family of exact, non-linear, solutions to the
one-dimensional incompressible MHD equations in the shearing sheet
limit.  As noted in \citet{GX94}, the only non-linear terms, which are
present through the perturbed magnetic energy density, are irrelevant
in the incompressible case under study (i.e., the total pressure can
be found {\it a posteriori} using the condition $\del \bcdot
\bb{v}=0$).

Due to the divergenceless nature of the disturbances under
consideration, the fluctuations in the vertical coordinate, $\delta
v_z(z)$ and $\delta B_z(z)$, reduce to a constant. Without loss of
generality, we take both constants to be zero\footnote{Note that the
fluctuations studied in \citet{GX94} are a particular case of the more
general solutions (\ref{eq:mean_plus_fluctuations_v}) and
(\ref{eq:mean_plus_fluctuations_b}).}.  We can further simplify the
system of equations (\ref{eq:euler}) and (\ref{eq:induction}) by
removing the background shear flow $\bb{v}_{\rm shear} = - q \Omega_0
(r-r_0) \check{\bb{\phi}}$. We obtain the following set of equations
for the fluctuations
\begin{eqnarray}
\label{eq:vx}
\frac{\partial}{\partial t} \delta v_r
&=& 2 \Omega_0 \delta v_\phi + \frac{\bar B_z}{4\pi\rho} \,
\frac{\partial}{\partial z} \delta B_r\,,  \\
\label{eq:vy}
\frac{\partial}{\partial t} \delta v_\phi 
&=& - (2-q)\Omega_0 \delta v_r + \frac{\bar B_z}{4\pi\rho}  \,
\frac{\partial}{\partial z} \delta B_\phi \,, \\
\label{eq:bx}
\frac{\partial}{\partial t} \delta B_r &=&  \bar B_z 
\frac{\partial}{\partial z} \delta v_r \,,  \\
\label{eq:by}
\frac{\partial}{\partial t} \delta B_\phi &=& - q \Omega_0 \delta B_r + 
\bar B_z \frac{\partial}{\partial z} \delta v_\phi \,,
\end{eqnarray}
where the first term on the right hand side of equation (\ref{eq:vy})
is related to the epicyclic frequency $\kappa\equiv\sqrt{2(2-q)}\,
\Omega_0$, at which the flow variables oscillate in an unmagnetized
disc.

It is convenient to define the new variables $\delta b_i \equiv \delta
B_i/\sqrt{4\pi\rho}$ for $i=r,\phi$, and introduce dimensionless
quantities by considering the characteristic time- and length-scales
set by $1/\Omega_0$ and $\bar{B}_z/(\sqrt{4\pi\rho}\,\Omega_0)$.  The
equations satisfied by the dimensionless fluctuations, $\delta
\tilde{v}_i$, $\delta \tilde{b}_i$, are then given by
\begin{eqnarray}
\label{eq:vx_nodim}
\partial_{\tilde t} \delta \tilde{v}_r &=& 2 \delta \tilde{v}_\phi + 
\partial_{\tilde z} \delta \tilde{b}_r \,,  \\
\label{eq:vy_nodim}
\partial_{\tilde t} \delta \tilde{v}_\phi &=& - (2-q) \delta \tilde{v}_r + 
\partial_{\tilde z} \delta \tilde{b}_\phi \,, \\
\label{eq:bx_nodim}
\partial_{\tilde t} \delta \tilde{b}_r &=&  
\partial_{\tilde z} \delta \tilde{v}_r \,,  \\
\label{eq:by_nodim}
\partial_{\tilde t} \delta \tilde{b}_\phi &=& - q \delta \tilde{b}_r + 
\partial_{\tilde z} \delta \tilde{v}_\phi \,,
\end{eqnarray}
where $\tilde t$ and $\tilde z$ denote the dimensionless time and
vertical coordinate, respectively. 

In order to simplify the notation, we drop hereafter the tilde
denoting the dimensionless quantities.  In the rest of the paper, all
the variables are to be regarded as dimensionless, unless otherwise
specified.

\section{The Eigenvalue Problem for the MRI\,: A Formal Solution}

In this section we provide a complete solution to the set of equations
(\ref{eq:vx_nodim})--(\ref{eq:by_nodim}) in Fourier space.  Taking the
Fourier transform of this set with respect to the $z$-coordinate, we
obtain the matrix equation
\begin{equation}
\partial_t \hat{\bb{\delta}}(k_n, t) = L  \, \hat{\bb{\delta}}(k_n,t) \,,
\label{eq:matrix_form} 
\end{equation}
where the vector $\hat{\bb{\delta}}(k_n, t)$ stands for 
\begin{equation}
\hat{\bb{\delta}}(k_n, t) = 
  \left[\begin{array}{c}
    \hat{\delta v_r}(k_n,t) \\ \hat{\delta v_\phi}(k_n,t) \\ 
    \hat{\delta b_r}(k_n,t) \\ \hat{\delta b_\phi}(k_n,t)
  \end{array}\right] 
\end{equation}
and $L$ represents the matrix
\begin{equation}
L = 
\left[\begin{array}{cccc}
    0 & 2  & i k_n  & 0 \\
    -(2-q)  & 0 & 0 & i k_n  \\
    i k_n  & 0 & 0 & 0 \\
    0 & i k_n  & -q  & 0
  \end{array}\right] \,. 
\end{equation}
The functions denoted by $\hat f(k_n, t)$ correspond to the Fourier
transform of the real functions, $f(z,t)$, and are defined via
\begin{equation}
\label{eq:ft_discrete}
\hat f(k_n,t) \equiv \frac{1}{2H}\int_{-H}^{H} f(z,t) \, e^{-ik_nz}
\,dz \,,
\end{equation}
where $H$ is the (dimensionless) scale-height and
$k_n$ is the wavenumber in the $z$-coordinate,
\begin{equation}
\label{eq:invft_discrete}
k_n \equiv \frac{n\pi}{H} \,,
\end{equation}
with $n$ being an integer number. Here, we have assumed periodic
boundary conditions at $z=\pm H$.

In order to solve the matrix equation (\ref{eq:matrix_form}), it is
convenient to find the base of eigenvectors, $\{\mathbf{e}_j\}$ with
$j=1,2,3,4$, in which $L$ is diagonal. This basis exists for all
values of the wavenumber $k_n$ (i.e., the rank of the matrix $L$ is
equal to 4, the dimension of the complex space) except for $k_n=0$ and
$k_n = \sqrt{2q}$.  In this base, the action of $L$ over the set
$\{\mathbf{e}_j\}$ is equivalent to a scalar multiplication, i.e.,
\begin{equation}
L_{\rm diag} \, \mathbf{e}_{j} = \sigma_j \,\mathbf{e}_{j}  \quad
\textrm{for} \quad j=1,2,3,4 \,,
\end{equation}
where $\{\sigma_j\}$ are complex scalars.

\subsection{Eigenvectors}

In the base of eigenvectors, the matrix $L$ has a diagonal
representation $L_{\rm diag}$ = diag$(\sigma_1, \sigma_2, \sigma_3,
\sigma_4)$. The eigenvalues $\{\sigma_j\}$, with $j=1,2,3,4$, are the
roots of the characteristic polynomial associated with $L$, i.e., the
dispersion relation associated with the MRI \citep{BH91,BH98},
\begin{equation}
\label{eq:dispersion_relation}
(k_n^2+\sigma_j^2)^2 + 
2(2 - q) ( k_n^2  + \sigma_j^2) - 4 k_n^2   = 0 \,,
\end{equation}
and are given by
\begin{equation}
\sigma_j = \pm \left(- \Lambda \pm \sqrt{\Delta}\,\right)^{1/2} \,,
\end{equation}
where we have defined the quantities $\Lambda$ and $\Delta$ such that
\begin{eqnarray}
  \Lambda &\equiv& k_n^2  + (2 - q)  \,, \\ \Delta &\equiv&
  (2- q)^2 + 4 k_n^2   \,.
\end{eqnarray}

For the modes with wavenumbers smaller than
\begin{equation}
\label{eq:k_BH}
k_{\rm BH} \equiv \sqrt{2 q} \,,
\end{equation}
i.e., the largest unstable wavenumber for the the MRI, the difference
$\sqrt{\Delta}-\Lambda$ is positive and we can define the ``growth
rate'' $\gamma$ and the ``oscillation frequency'' $\omega$ by
\begin{eqnarray}
\label{eq:gamma}
\gamma   &\equiv& \left(\sqrt{\Delta} -\Lambda\right)^{1/2} \,, \\
\label{eq:omega}
 \omega &\equiv& \left(\sqrt{\Delta} + \Lambda\right)^{1/2} \,,
\end{eqnarray}
both of which are real and positive (for all positive values of the
parameter $q$). This shows that two of the solutions of equation
(\ref{eq:dispersion_relation}) are real and the other two are
imaginary.  We can thus write the four eigenvalues in compact notation
as
\begin{eqnarray}
\label{eq:eigenvalues}
\sigma_1 = \gamma \,, \quad
\sigma_2 = -\gamma \,,\quad
\sigma_3 = i\omega \,, \quad \textrm{and} \quad
\sigma_4 = -i\omega \,.
\end{eqnarray}

The set of normalized eigenvectors, $\{\mathbf{e}_{\sigma_j}\}$,
associated with these eigenvalues can be written as 
\begin{eqnarray}
\label{eq:eigenvectors}
\mathbf{e}_{\sigma_j} &\equiv& \frac{\mathbf{e}_{j}}{\|\mathbf{e}_{j}\|}
\quad \textrm{for} \quad j=1,2,3,4 \,,
\end{eqnarray}
where
\begin{equation}
\label{eq:e_sigma_j}
\mathbf{e}_{j}(k_n) =
    \left[\begin{array}{c}
      \sigma_j \\
      (k_n^2  + \sigma_j^2)/2 \\
      i k_n  \\
      - i 2 k_n  \sigma_j /(k_n^2  + \sigma_j^2)
    \end{array}\right],
\end{equation}
and the norms are given by
\begin{equation}
\|\mathbf{e}_{j}\| 
\equiv \left[\sum_{l=1}^{4} 
{\rm e}^{l}_{j} 
{\rm e}^{l*}_{j}\right]^{1/2} \,,
\end{equation}
where ${\rm e}^{l}_{j}$ is the $l$-th component of the (unnormalized)
eigenvector associated with the eigenvalue $\sigma_j$.  This set of
four eigenvectors $\{\mathbf{e}_{\sigma_j}\}$, together with the set
of scalars $\{\sigma_j\}$, constitute the full solution to the
eigenvalue problem defined by the MRI.

The roots $\sigma_j$ of the characteristic polynomial are not
degenerate. Because of this, the set $\{\mathbf{e}_{\sigma_j}\}$
constitutes a basis set of four independent (complex) vectors that are
able to span ${\mathbb C}^4$, i.e., the space of tetra-dimensional
complex vectors, for each value of $k_n$, provided that $\gamma$ and
$\omega$ are given by equations (\ref{eq:gamma}) and (\ref{eq:omega}),
respectively. Note, however, that they will not in general be
orthogonal, i.e., $\mathbf{e}_{\sigma_j} \bcdot \,
\mathbf{e}_{\sigma_{j'}} \ne 0$ for $j \ne j'$.  If desired, an
orthogonal basis of eigenvectors can be constructed using the
Gram-Schmidt orthogonalization procedure \citep[see, e.g.,][]{HK71}.

\subsection{Properties of the Eigenvectors}

\begin{figure}
  \includegraphics[width=\columnwidth,trim=0 20 0 20]{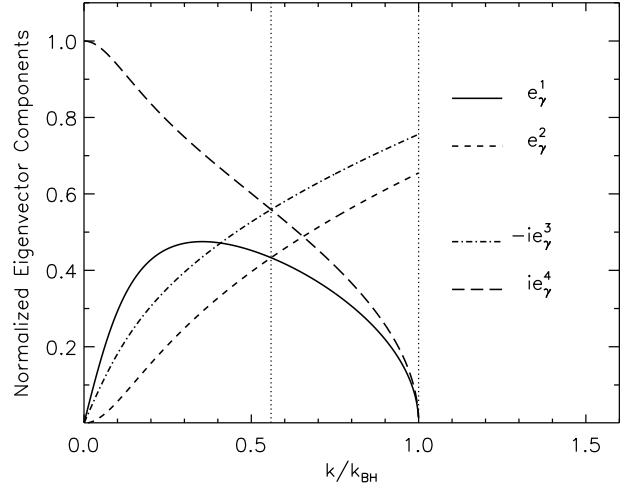}
  \caption{The components of the normalized unstable eigenvector
    $\mathbf{e}_{\gamma}$ defined in equation (\ref{eq:eigenvectors}).
    The vertical {\it dotted} lines denote the wavenumber
    corresponding to the most unstable mode, $k_{\rm max}$,
    (eq.\,[\ref{eq:k_max}]), and the largest unstable wavenumber,
    $k_{\rm BH}$, (eq.\,[\ref{eq:k_BH}]).}
  \label{fig:mri-fourier-amplitudes}
\end{figure}

Despite the complicated functional dependence of the various complex
eigenvector components on the wavenumber, $k_n$, some simple and
useful relations hold for the most relevant (unstable) eigenvector.
Figure \ref{fig:mri-fourier-amplitudes} shows the four components of
the eigenvector $\mathbf{e}_{\gamma}$ as a function of the wavenumber
for a Keplerian profile ($q=3/2$) and illustrates the fact that the
modulus of the different components satisfy two simple inequalities
\begin{eqnarray}
\label{eq:ineq_e_4gr1}
|{\rm e}^{4}_{\gamma}| \!\!\!& > & \!\!\!\!|{\rm e}^{1}_{\gamma}| \,, \\
\label{eq:ineq_e_3gr2}
|{\rm e}^{3}_{\gamma}| \!\!\!& > & \!\!\!\!|{\rm e}^{2}_{\gamma}| \,,
\end{eqnarray}
for all values of $0 < k_{n} < k_{\rm BH}$. These inequalities
do indeed hold for all values of the shear parameter $0<q<2$.

We also note that equation (\ref{eq:e_sigma_j}) exposes a relationship among
the components of any given eigenvector
$\mathbf{e}_{\sigma_j}$. It is immediate to see that
\begin{equation}
\label{eq:eigenvector_ratios}
\frac{-{\rm e}^{4}_{\sigma_j}}{{\rm e}^{1}_{\sigma_j}} = 
\frac{{\rm e}^{3}_{\sigma_j}}{{\rm e}^{2}_{\sigma_j}}  = 
\frac{2ik_n}{k_n^2 + \sigma_j^2} \,, \quad \textrm{for} \quad
j=1,2,3,4 \,.
\end{equation}
In particular, the following equalities hold for the components of the
unstable eigenvector, $\mathbf{e}_{\gamma}$,  at the wavenumber 
\begin{equation}
\label{eq:k_max}
k_{\rm max} \equiv  \frac{q}{2} \sqrt{\frac{4}{q}-1} \,,
\end{equation}
for which the growth rate is maximum, $\gamma_{\rm max} \equiv q/2$,
\begin{eqnarray}
\label{eq:e_1_2_at_kmax}
{\rm e}^{1}_{\gamma}(k_{\rm max}) & = \, \, \, \,
{\rm e}^{2}_{\gamma}(k_{\rm max}) & =  \,
\frac{1}{2} \, \sqrt{\frac{q}{2}} \,, \\
\label{eq:e_3_4_at_kmax}
{\rm e}^{3}_{\gamma}(k_{\rm max}) & = \,
-{\rm e}^{4}_{\gamma}(k_{\rm max})& =  \,
\frac{i}{2} \, \sqrt{2 -\frac{q}{2}}\,.
\end{eqnarray}

As we show in the next section, the inequalities
(\ref{eq:ineq_e_4gr1}) and (\ref{eq:ineq_e_3gr2}), together with
equations (\ref{eq:e_1_2_at_kmax}) and (\ref{eq:e_3_4_at_kmax}), play an
important role in establishing the relative magnitude of the different
mean stress components and mean magnetic energies associated with the
fluctuations in the velocity and magnetic fields (see also Appendix
\ref{sec:appendix_b}).

Finally, we stress here that the phase differences among the different
eigenvector components cannot be eliminated by a linear (real or
complex) transformation. In other words, it is not possible to obtain
a set of four real (or purely imaginary, for that matter) linearly
independent set of eigenvectors that will also form a basis in which
to expand the general solution to equation (\ref{eq:matrix_form}).
Taking into consideration the complex nature of the eigenvalue problem
in MRI is crucial when writing the physical solutions for the
spatio-temporal evolution of the velocity and magnetic field
fluctuations in terms of complex eigenvectors.  As we discuss in the
next section, this in turn has a direct implication for the
expressions that are needed in the calculation of the mean stresses in
physical (as opposed to spectral) space.

\subsection{Temporal Evolution}

We have now all the elements to solve equation (\ref{eq:matrix_form}).
In the base defined by $\{\mathbf{e}_{\sigma_j}\}$, any given vector
$\hat{\bb{\delta}}(k_n,t)$ can be written as
\begin{equation}
\hat{\bb{\delta}}(k_n,t) = \sum_{j=1}^{4} a_j(k_n,t) \,
\mathbf{e}_{\sigma_j} \,,
\end{equation}
where the coefficients $a_j(k_n,t)$, i.e., the coordinates
of  $\hat{\bb{\delta}}(k_n,t)$ in the eigenvector basis,
are the components of the vector $\bb{a}(k_n,t)$ obtained from 
the transformation
\begin{equation}
\bb{a}(k_n,t) = Q^{-1} \, \hat{\bb{\delta}}(k_n,t) \,.
\end{equation}
The matrix $Q^{-1}$ is the matrix for the change of coordinates from the
standard basis to the normalized eigenvector basis and can be obtained
by calculating the inverse of the matrix 
\begin{equation}
Q = [
\mathbf{e}_{\sigma_1} \ \ \mathbf{e}_{\sigma_2} \ \   
\mathbf{e}_{\sigma_3} \ \ \mathbf{e}_{\sigma_4}] \,.
\end{equation}

Multiplying equation (\ref{eq:matrix_form}) at the left side by 
$Q^{-1}$ and using the fact that 
\begin{equation}
L_\mathrm{diag}=Q^{-1}\,L\,Q \,,
\end{equation}
we obtain a matrix equation for the vector $\bb{a}(k_n,t)$,
\begin{equation}
\partial_t \bb{a}(k_n,t) = L_{\rm diag} \, \bb{a}(k_n,t) \,,
\end{equation}
which can be written in components as
\begin{equation}
\partial_t a_j(k_n,t) = \sigma_j \, a_j(k_n,t)  
\quad \textrm{with} \quad j=1,2,3,4 \,.
\end{equation}
The solution of these equations is then given by
\begin{equation}
a_j(k_n,t) = a_j(k_n,0) \, e^{\sigma_j t} \quad \textrm{with} \quad
j=1,2,3,4 \,.
\end{equation}

We can finally write the solution to equation
(\ref{eq:matrix_form}) as
\begin{equation}
\label{eq:solution_1}
\hat{\bb{\delta}}(k_n,t) = \sum_{j=1}^{4}  a_j(k_n,0) \,
e^{\sigma_j t} \,\mathbf{e}_{\sigma_j}  \,,
\end{equation}
where $\{\sigma_j\}$ and $\{\mathbf{e}_{\sigma_j}\}$, for $j=1,2,3,4$,
are given by equations
(\ref{eq:eigenvalues})
and
(\ref{eq:eigenvectors}), and the initial conditions
$\bb{a}(k_n,0)$ are related to the initial spectrum of fluctuations,
$\hat{\bb{\delta}}(k_n,0)$, via
$\bb{a}(k_n,0) = Q^{-1} \, \hat{\bb{\delta}}(k_n,0)$.

\section{Net Angular Momentum Transport by the MRI}
\label{sec:angular_momentum_transport}

Having obtained the solution for the temporal evolution of the velocity
and magnetic field fluctuations in Fourier space we can now explore
the effect of the MRI on the mean values of the Reynolds
and Maxwell stresses.

\subsection{Definitions}

The average over the disc scale-height, $2H$,
of the product of any two physical quantities, $f(z,t)\,g(z,t)$,
\begin{equation}
\label{eq:f_mean}
\overline{f g}\,(t) \equiv \frac{1}{2H}\int_{-H}^{H} f(z,t)\, g(t,z)\,dz \,,
\end{equation}
can be written in terms of their corresponding Fourier transforms,
$\hat f(k_n,t)$ and $\hat g(k_n,t)$, as (see Appendix \ref{sec:appendix_a})
\begin{eqnarray}
\label{eq:mean_ft}
\overline{f g}\,(t) =  2 \sum_{n=1}^{\infty} 
\; {\rm Re}[\,\hat{f}(k_n,t)\, \hat{g}^*(k_n,t)\,] \,.
\end{eqnarray}
Here, ${\rm Re}[\,\,]$ stands for the real part of the quantity
between brackets, the asterisk in $\hat{g}^*(k_n,t)$ denotes the
complex conjugate, and we have considered that the functions $f(z,t)$,
and $g(z,t)$ (both with zero mean) are real and, therefore, their
Fourier transforms satisfy $\hat{f}(-k_n,t) = \hat{f}^*(k_n, t)$.

Using equation (\ref{eq:mean_ft}), we can write the mean values of the
quantities
\begin{eqnarray}
  R_{ij}(z,t) & \equiv &  \delta v_i(z,t) \, \delta v_j(z,t) \, , \\ 
  M_{ij}(z,t) & \equiv &  \delta b_i(z,t) \, \delta b_j(z,t) \, ,
\end{eqnarray}
with $i, j=r,\phi$,  as
\begin{eqnarray}
\label{eq:mean_R}
\bar{R}_{ij}(t) & \equiv & 2 \sum_{n=1}^{\infty} \;
\real[\,
\hat{\delta v_i}(k_n,t) \, \hat{\delta v_j}\!\!^*\!(k_n,t)\,]
\,, \\
\label{eq:mean_M}
\bar{M}_{ij}(t) & \equiv & 2 \sum_{n=1}^{\infty} \;
\real[\,
\hat{\delta b_i}(k_n,t) \, \hat{\delta b_j}\!\!^*\!(k_n,t)\,] \,, 
\end{eqnarray}
where the temporal evolution of the fluctuations in Fourier space,
$\hat{\delta v_r}(k_n,t)$,
$\hat{\delta v_\phi}(k_n,t)$,
$\hat{\delta b_r}(k_n,t)$, and
$\hat{\delta b_\phi}(k_n,t)$,
is governed by equation (\ref{eq:solution_1}).

\subsection{Properties of the MRI-Driven Stresses}

At late times, during the exponential growth of the instability, the
branch of unstable modes will dominate the growth of the fluctuations
and we can write the most important (secular) contribution to the mean
stresses by defining
\begin{eqnarray}
\label{eq:mean_R_perk}
\bar{R}_{r\phi}(t) & = & 2 \sum_{n=1}^{N_\mathrm{BH}}
\mathcal{R}_{r\phi}(k_n) 
\,|a_1|^2 e^{ 2\gamma t} + \dots \,,\\
\label{eq:mean_M_perk}
\bar{M}_{r\phi}(t) & = & 2 \sum_{n=1}^{N_\mathrm{BH}}
\mathcal{M}_{r\phi}(k_n) 
\,|a_1|^2 e^{ 2\gamma t}  + \dots \,,
\end{eqnarray}
where $N_{\rm BH}$ is the index associated with the largest unstable
wavenumber (i.e., the mode labelled with highest $k_n<k_{\rm
BH}$)\footnote{Extending the summations to include the non-growing
modes with $k_n>k_{\rm BH}$ would only add a negligible oscillatory
contribution to the mean stresses.}, the dots represent terms that
grow at most as fast as $e^{\gamma t}$, and we have defined the
functions
\begin{equation}
\label{eq:R_perk}
\mathcal{R}_{r\phi}(k_n) = 
\frac{\real[\textrm{e}_\gamma^1 \, \textrm{e}_\gamma^{2*}]}
{\|\mathbf{e}_{1}\|^2} \,,
\end{equation}
and
\begin{equation}
\label{eq:M_perk}
\mathcal{M}_{r\phi}(k_n) =
\frac{\real[\textrm{e}_\gamma^3 \, \textrm{e}_\gamma^{4*}]}
{\|\mathbf{e}_{1}\|^2} \,.
\end{equation}
Note that these functions are not the Fourier transforms of the
Reynolds and Maxwell stresses, but rather represent the contribution
of the fluctuations at the scale $k_n$ to the corresponding mean
physical stresses.  We will refer to these quantities as the {\it
per}-$k$ contributions to the mean.  

\begin{figure}
\includegraphics[width=\columnwidth,trim=0 20 0 20]{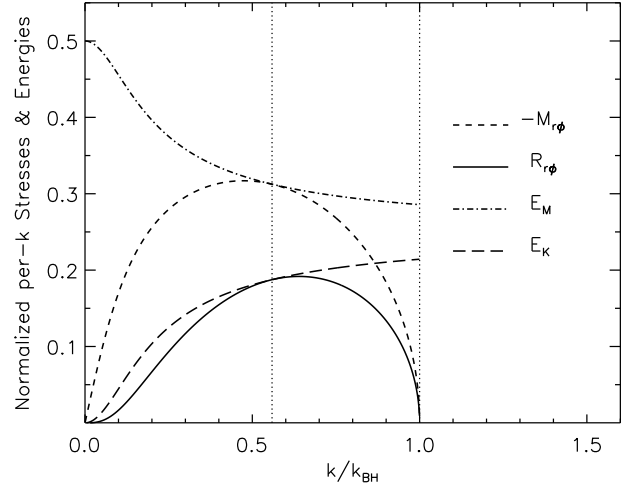}
\caption{The {\it per}-$k$ contributions, $\mathcal{R}_{r\phi}(k_n)$,
$\mathcal{M}_{r\phi}(k_n)$, $\mathcal{E}_{K}(k_n)$, and
$\mathcal{E}_{M}(k_n)$, associated with the corresponding mean
physical stresses (\ref{eq:mean_R_perk}) and (\ref{eq:mean_M_perk})
and mean physical magnetic and kinetic energy densities, defined in
Appendix \ref{sec:appendix_b}, eqs.\,(\ref{eq:mean_EK_perk}) and
(\ref{eq:mean_EM_perk}).  The vertical {\it dotted} lines denote the
wavenumber corresponding to the most unstable mode, $k_{\rm max}$,
(eq.\,[\ref{eq:k_max}]), and the largest unstable wavenumber, $k_{\rm
BH}$, (eq.\,[\ref{eq:k_BH}]).}
\label{fig:mri-stresses}
\end{figure}

The complex nature of the various components of the unstable
eigenvector, together with the inequalities
(\ref{eq:ineq_e_4gr1}) and (\ref{eq:ineq_e_3gr2}), dictate the relative
magnitude of the {\it per}-$k$ contributions associated with the
Maxwell and Reynolds stresses and the magnetic and kinetic energy
densities (see Appendix \ref{sec:appendix_b} for a detailed discussion
of the latter case).  Figure \ref{fig:mri-stresses} shows the
functions $\mathcal{R}_{r\phi}(k_n)$ and $\mathcal{M}_{r\phi}(k_n)$
for a Keplerian profile ($q=3/2$).  It is evident from this figure
that, in this case, the {\it per}-$k$ contribution of the
Maxwell stress is always larger than the the {\it per}-$k$
contribution corresponding to the Reynolds stress, i.e., 
$-\mathcal{M}_{r\phi}(k_n) > \mathcal{R}_{r\phi}(k_n)$.  This is
indeed true for all values of the shear parameter $0<q<2$ (see below).

The coefficients $\textrm{e}_\gamma^j$ for $j=1,2,3,4$ in equations
(\ref{eq:R_perk}) and (\ref{eq:M_perk}) are the components of the
(normalized) unstable eigenvector given by equation
(\ref{eq:e_sigma_j}) with $\sigma_1 = \gamma$. We can then write, the
mean values of the Reynolds and Maxwell stresses, to leading order in
time, as
\begin{eqnarray}
\label{eq:mean_Rrphi}
\bar{R}_{r\phi}(t) & = & \sum_{n=1}^{N_\mathrm{BH}}
\gamma \, (k_n^2  + \gamma^2) 
\, \frac{|a_1|^2}{\|\mathbf{e}_{1}\|^2} \, e^{ 2\gamma t} \,,\\
\label{eq:mean_Mrphi}
\bar{M}_{r\phi}(t) & = & - 4 \sum_{n=1}^{N_\mathrm{BH}}
\frac{\gamma \, k_n^2 }{k_n^2  + \gamma^2}  
\, \frac{|a_1|^2}{\|\mathbf{e}_{1}\|^2} \, e^{ 2\gamma t} \,.
\end{eqnarray}

Equations (\ref{eq:mean_Rrphi}) and (\ref{eq:mean_Mrphi}) show
explicitly that the mean Reynolds and Maxwell stresses will be,
respectively, positive and negative,
\begin{equation}
\bar{R}_{r\phi}(t) >0 \qquad \textrm{and} \qquad \bar{M}_{r\phi}(t) < 0 \,.
\end{equation}
This, in turn, implies that the mean total MRI-driven stress will be
always positive, i.e.,
\begin{equation}
\bar{T}_{r\phi}(t) = \bar{R}_{r\phi}(t) - \bar{M}_{r\phi}(t)  >0 \,,
\end{equation}
driving a net outward flux of angular momentum as discussed in
\S\ref{sec:intro}.

It is not hard to show now that the magnitude of the Maxwell stress,
$-\bar{M}_{r\phi}(t)$, will always be larger than the magnitude
of the Reynolds stress, $\bar{R}_{r\phi}(t)$, provided that the shear
parameter is $q<2$.  In order to see that this is the case, it is
enough to show that the ratio of the {\it per}-$k$ contributions to
the Reynolds and Maxwell stresses, defined in equations
(\ref{eq:R_perk}) and (\ref{eq:M_perk}), satisfy
\begin{equation}
\label{eq:Mrphi_Rrphi_ratio}
\frac{-\mathcal{M}_{r\phi}(k_n)} {\,\,\mathcal{R}_{r\phi}(k_n)} =
\frac{4 k_n^2 }{(k_n^2 + \gamma^2)^2} > 1 \,,
\end{equation}
for all the wavenumbers $k_n$.
Adding and subtracting the factor $(k_n^2  + \gamma^2)^2$
in the numerator and using the dispersion relation
(\ref{eq:dispersion_relation}) we obtain,
\begin{equation}
\frac{-\mathcal{M}_{r\phi}(k_n)}{\mathcal{R}_{r\phi}(k_n)} =
1 + \frac{2(2-q)}{k_n^2  + \gamma^2} \,,
\end{equation}
which is clearly larger than unity for all values of $k_n$ provided
that $q<2$.  It is then evident  that the mean Maxwell
stress will be larger than the mean Reynolds stress as long as
the flow is Rayleigh-stable, i.e.,
\begin{equation}
-\bar{M}_{r\phi}(t) > \bar{R}_{r\phi}(t)  \quad \textrm{for} \quad 0<q<2 \,.
\end{equation}
This inequality provides analytical support to the results obtained in
numerical simulations, i.e., that the Maxwell stress constitutes the
major contribution to the total stress in magnetized accretion discs
\citep[see, e.g.,][]{HGB95}.

We conclude this section by calculating the ratio
$-\bar{M}_{r\phi}(t)/\bar{R}_{r\phi}(t)$ at late times
during the exponential growth of the instability.  For times
that are long compared to the dynamical time-scale, the unstable mode
with maximum growth dominates the dynamics of the mean stresses and
the sums over all wavenumbers in equations (\ref{eq:mean_Rrphi}) and
(\ref{eq:mean_Mrphi}) can be approximated by a single term
corresponding to $k_n=k_{\rm max}$. In this case, we can use equations
(\ref{eq:e_1_2_at_kmax}) and (\ref{eq:e_3_4_at_kmax}) to write
\begin{equation}
\lim_{t\gg1}
\label{eq:analytical_ratio}
\frac{-\bar{M}_{r\phi}(t)}{\,\,\bar{R}_{r\phi}(t)} = -
\left.\frac
{\real[\textrm{e}_\gamma^{3} \textrm{e}_\gamma^{4*}]\,}
{\real[\textrm{e}_\gamma^{1} \textrm{e}_\gamma^{2*}]\,} \,
\right|_{k_{\rm max}} \!\!\!  = \frac{4-q}{q} \,.
\end{equation}
This result shows explicitly that the ratio
$-\bar{M}_{r\phi}(t)/\bar{R}_{r\phi}(t)$ depends only on the
shear parameter and not on the magnitude or even the sign of the
magnetic field. These same conclusions can be drawn for the ratio between
the magnetic and kinetic energies associated with the fluctuations
(see Appendix \ref{sec:appendix_b} for a detailed discussion).

\section{Discussion}

In this paper we have studied the properties of the mean Maxwell and
Reynolds stresses in a differentially rotating flow during the
exponential growth of the magnetorotational instability and have
identified its signature in their temporal evolution.  In order to
achieve this goal, we obtained the complex eigenvectors associated
with the magnetorotational instability and presented the formalism
needed to calculate the temporal evolution of the mean Maxwell and
Reynolds stresses in terms of them.

We showed that, during the phase of exponential growth characterizing
the instability, the mean values of the Reynolds and Maxwell stresses
are always positive and negative, respectively, i.e.,
$\bar{R}_{r\phi}(t) > 0$ and $\bar{M}_{r\phi}(t) < 0$.  This leads,
automatically, to a net outward angular momentum flux mediated
by a total mean positive stress, $\bar{T}_{r\phi}(t) =
\bar{R}_{r\phi}(t) - \bar{M}_{r\phi}(t) > 0$.  We further demonstrated
that, for a flow that is Rayleigh-stable (i.e., when
$q=-d\ln\Omega/d\ln r<2$), the contributions to the total stress
associated with the correlated magnetic fluctuations are always larger
than the contributions due to the correlations in the velocity
fluctuations, i.e., $-\bar{M}_{r\phi}(t)>\bar{R}_{r\phi}(t)$.

We also proved that, during the late times of the linear phase of the
instability, the ratio of the Maxwell to the Reynolds stresses simply
becomes
\begin{equation}
\label{eq:analytical_ratio_late}
\lim_{t\gg1}
\frac{-\bar{M}_{r\phi}(t)}{\,\,\bar{R}_{r\phi}(t)} =
\frac{4-q}{q} \,.
\end{equation}
This is a remarkable result, because it does not depend on the initial
spectrum of fluctuations or the value of the seed magnetic field. It
is, therefore, plausible that, even in the saturated state of the
instability, when fully developed turbulence is present, the ratio of
the Maxwell to the Reynolds stresses has also a very weak dependence
on the properties of the turbulence and is determined mainly by the
local shear.

\begin{figure}
\includegraphics[width=\columnwidth,trim=0 20 0 20]{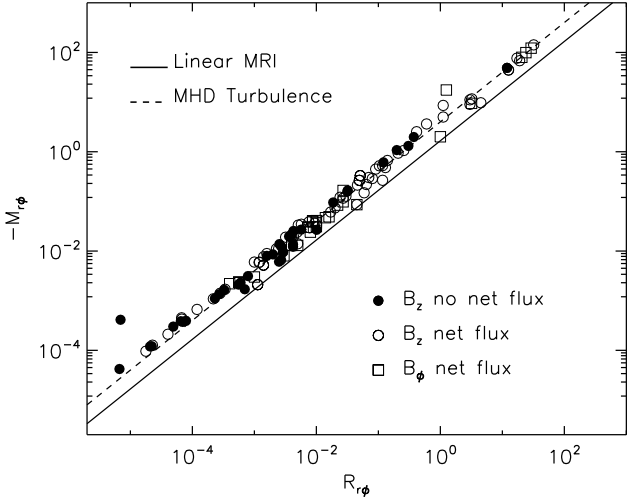}
\caption{The correlation between the Maxwell stress,
$\bar{M}_{r\phi}$, and the Reynolds stress, $\bar{R}_{r\phi}$, at
saturation obtained in numerical simulations of shearing boxes with
Keplerian velocity profiles but different initial conditions. {\it
Filled circles} correspond to simulations with zero net vertical
magnetic flux. {\it Open circles} correspond to simulations with
finite net vertical magnetic flux. {\it Squares} correspond to
simulations with an initial toroidal field. (See the text for
references.) The {\it dashed line} is the correlation often quoted in
the literature.  The {\it solid line} corresponds to the late-time
ratio of the two stresses during the exponential growth of the MRI, as
predicted by equation~(\ref{eq:analytical_ratio_late}) for $q=3/2$.}
\label{fig:mrf_vs_rrf}
\end{figure}

For shearing box simulations with a Keplerian velocity profile, the
ratio of the Maxwell to the Reynolds stresses in the saturated state
has been often quoted to be constant indeed (of order $\simeq 4$),
almost independent of the setup of the simulation, the initial
conditions, and the boundary conditions. This is shown in
Figure~\ref{fig:mrf_vs_rrf}, where we plot the correlation between the
Maxwell stress, $\bar{M}_{r\phi}$, and the Reynolds stress,
$\bar{R}_{r\phi}$, at saturation, for a number of numerical
simulations of shearing boxes, with Keplerian velocity profiles but
different initial conditions (data points are from~\citealt{HGB95,
Stone96, Fleming03, Sanoetal04, GS05}).  It is remarkable that this
linear correlation between the stresses, in fully developed turbulent
states resulting from very different sets of initial conditions, spans
over six orders of magnitude.  The ratio
$-\bar{M}_{r\phi}/\bar{R}_{r\phi} = 4$ is shown in the same figure
with a {\it dashed line}, whereas the {\it solid line} shows the ratio
obtained from equation~(\ref{eq:analytical_ratio_late}) for $q=3/2$,
i.e., $-\bar{M}_{r\phi}/\bar{R}_{r\phi}=5/3$, characterizing the
linear phase of the instability. Hence, to within factors of order
unity, the ratio between the stresses in the turbulent state seems to
be independent of the initial set of conditions over a wide range of
parameter space and to be similar to the value set during the linear
phase of the instability.

\begin{figure}
\includegraphics[width=\columnwidth,trim=0 20 0 20]{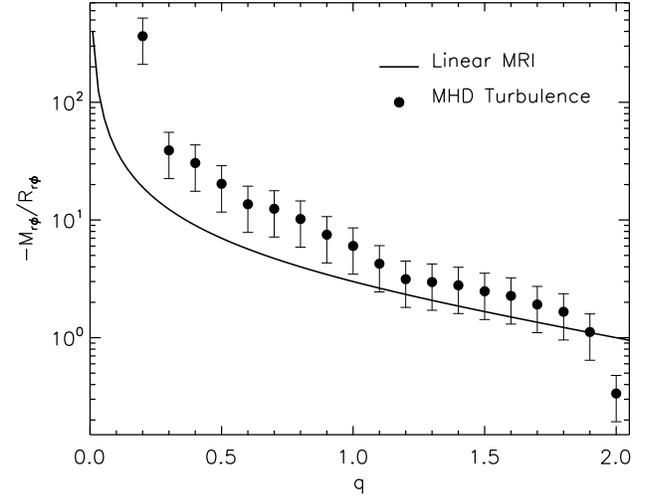}
\caption{The dependence of the ratio of the mean Maxwell to the mean
Reynolds stresses, $-\bar{M}_{r\phi}/\bar{R}_{r\phi}$, on the shear
parameter $q$. The data points correspond to the results of shearing
box simulations by~\citet{HBW99} in the saturated state.  The {\it
solid line} shows the analytic result
(eq.\,[\ref{eq:analytical_ratio_late}]) for the ratio of the mean
stresses during the late time of the exponential growth of the MRI.}
\label{fig:analytical-vs-simulations}
\end{figure}

The dependence of the ratio of the stresses
on the shear parameter, $q$, has not been studied extensively with
numerical simulations so far. The only comprehensive study is
by~\citet{HBW99} and their result is shown in
Figure~\ref{fig:analytical-vs-simulations}
\footnote{Note that, \citet{HBW99} quote the average stresses and the
width of their distribution throughout the simulations, but not the
uncertainty in the mean values.  In
Fig.~\ref{fig:analytical-vs-simulations}, we have assigned a nominal
30\% uncertainty to their quoted mean values.  This is comparable to
the usual quoted uncertainty for the stresses and is also comparable
to the spread in Fig.~\ref{fig:mrf_vs_rrf}.}.  Superimposed on the
figure is the analytic prediction, equation
(\ref{eq:analytical_ratio_late}), for the ratio of the stresses as a
function of the shear $q$ at late times during the exponential growth
of the MRI.  In this case, the qualitative trend followed by the ratio
of the stresses at saturation as a function of the shear parameter $q$
seems also to be similar to that obtained at late times during the
linear phase of the instability.

The simultaneous analysis of Figures \ref{fig:mrf_vs_rrf} and
\ref{fig:analytical-vs-simulations} demonstrates that the ratio
$-\bar{M}_{r\phi}/\bar{R}_{r\phi}$ during the turbulent saturated
state in local simulations of accretion discs is determined almost
entirely by the local shear and depends very weakly on the other
properties of the flow or the initial conditions.  These figures also
show that the ratios of the Maxwell to the Reynolds stresses
calculated during the turbulent saturated state are qualitatively
similar to the corresponding ratios found during the late times of the
linear phase of the instability, even though the latter are slightly
lower (typically by a factor of 2).  This is remarkable because, when
deriving equation~(\ref{eq:analytical_ratio_late}) we have assumed
that the MHD fluid is incompressible and considered only fluctuations
that depend on the vertical, $z$, coordinate. Moreover, in the spirit
of the linear analysis, we have not incorporated energy cascades
between different scales, neither did we consider dissipation or
reconnection processes that lead to saturation.  Of course, the
numerical simulations addressing the non-linear regime of the
instability do not suffer from any of the approximations invoked to
solve for the temporal evolution of the stresses during the phase of
exponential growth. Nevertheless, the ratio of the Maxwell to the
Reynolds stresses that characterize the turbulent saturated state are
similar (to within factors of order unity) to the ratios
characterizing the late times of the linear phase of the instability.

\section*{Acknowledgments}
We thank an anonymous referee for useful comments that contributed to
improving the clarity of this paper.  This work was partially
supported by NASA grant NAG-513374.

\appendix

\section{Mean Values of Correlation Functions}
\label{sec:appendix_a}

For convenience, we provide here a brief demonstration of equation
(\ref{eq:mean_ft}). The relationship between the mean value of the
product of two real functions, $\overline{fg}$, and their
corresponding Fourier transforms can be obtained by substituting the
expressions for $f(z,t)$ and $g(z,t)$ in terms of their Fourier
series, i.e.,
\begin{equation}
f(z,t) \equiv \sum_{n=-\infty}^{\infty} \hat f(k_n,t) \, e^{ik_nz}  \,,
\end{equation}
into the expression for the mean value,
\begin{equation}
\overline{f g}\,(t) \equiv \frac{1}{2H}\int_{-H}^{H} f(z,t)\, g(t,z)\,dz \,.
\end{equation}
The result is,
\begin{equation}
\overline{f g}\,(t) = \!\!\!
\sum_{n,m=-\infty}^{\infty} \!\!\! \hat f(k_n,t) \, \hat{g}^*(k_{m},t) \,
\frac{1}{2H}\int_{-H}^{H} \!\!\!   e^{i(k_n-k_{m})z} \, dz \,,
\end{equation}
which can be rewritten using the orthogonality of the Fourier
polynomials in the interval $[-H,H]$ as
\begin{equation}
\overline{f g}\,(t)= \sum_{n=-\infty}^{\infty} \hat f(k_n,t) \hat
g^*(k_n,t) \,.
\end{equation}
This is the discrete version of Plancherel's theorem which states that
the Fourier transform is an isometry, i.e., it preserves the inner
product \citep[see., e.g.,][]{Shilov73}.

Denoting the Fourier transforms $\hat f(k_n,t)$ and $\hat{g}^*(k_n,t)
$ by $\hat f_{n}$ and $\hat{g}^*_{n}$ in order to simplify the
notation, we can write the following series of identities
\begin{eqnarray}
\overline{f g}\,(t) 
&=& \sum_{n=-\infty}^{\infty} \hat f_n \hat g^*_n \,, \\
&=& \hat{f}_{0}\,\hat{g}_{0} 
+ \sum_{n=1}^{\infty} \hat{f}_{n} \hat{g}^*_{n}  +  
  \sum_{n= -\infty}^{-1} \hat{f}_{n} \hat{g}^*_{n} \\
&=& \hat{f}_{0}\,\hat{g}_{0} + 
\sum_{n=1}^{\infty} \left[\hat{f}_{n} \hat{g}^*_{n} + 
                          \hat{f}_{-n} \hat{g}^*_{-n}\right] \\
&=& \hat{f}_{0}\,\hat{g}_{0} + 
\sum_{n=1}^{\infty} \left[\hat{f}_{n} \hat{g}_{n}^*  + 
                          \hat{f}_{n}^* \hat{g}_{n}\right] \\
&=& \hat{f}_{0}\,\hat{g}_{0} + 
2 \sum_{n=1}^{\infty} \;\real[\hat{f}_{n} \hat{g}_{n}^*] \,,
\end{eqnarray}
where we have used the fact that the functions $f(z,t)$ and $g(z,t)$
are real and, hence, their Fourier transforms satisfy
$\hat{f}_{-n} = \hat{f}^*_n$ and $\hat{g}_{-n} = \hat{g}^*_n$.  Note
that the factors $\hat{f}_0$ and $\hat{g}_0$ are just the mean values
of the functions $f(z,t)$ and $g(z,t)$ and, therefore, do not
contribute to the final expression in equation (\ref{eq:mean_ft}).
It is then clear that, no matter whether the initial Fourier
transforms corresponding to the functions $f(z,t)$ and $g(z,t)$ are
real or imaginary, the mean value, $\overline{f g}\,(t)$, will be well
defined.

\section{Energetics of MRI-driven Fluctuations}
\label{sec:appendix_b}

The relationships given in equation (\ref{eq:eigenvector_ratios}) lead
to identities and inequalities involving the different mean stress
components and the mean kinetic and magnetic energies associated with
the fluctuations. In particular, the total mean stress is bounded by
the total mean energy of the fluctuations. Moreover, the ratio of the
mean magnetic to the mean kinetic energies is equal to the (absolute
value of the) ratio between the mean Maxwell and the mean Reynolds
stresses given by equation (\ref{eq:analytical_ratio_late}).

As we defined the mean stresses in terms of their {\it per}-$k$
contributions in \S \ref{sec:angular_momentum_transport}, we can also
define, to leading order in time, the mean energies associated with
the fluctuations in the velocity and magnetic field by
\begin{eqnarray}
\label{eq:mean_EM_perk}
\bar{E}_K(t) = 2 \sum_{n=1}^{k_\mathrm{BH}} \mathcal{E}_K(k_n)
\,|a_1|^2 e^{ 2\gamma t} \,, \\
\label{eq:mean_EK_perk}
\bar{E}_M(t) =   2 \sum_{n=1}^{k_\mathrm{BH}} \mathcal{E}_M(k_n) 
\,|a_1|^2 e^{ 2\gamma t} \,,
\end{eqnarray}
where the corresponding {\it per}-$k$ contributions are given by
\begin{eqnarray}
\mathcal{E}_K(k_n)  &=& \frac{1}{2} \,
[\mathcal{R}_{rr}(k_n) + \mathcal{R}_{\phi\phi}(k_n)] \,,\\
\mathcal{E}_M(k_n) &=& \frac{1}{2} \, [\mathcal{M}_{rr}(k_n) +
\mathcal{M}_{\phi\phi}(k_n)] \,.
\end{eqnarray}
Figure \ref{fig:mri-stresses} shows the dependences of the functions
$\mathcal{E}_{K}(k_n)$ and $\mathcal{E}_{M}(k_n)$ for a Keplerian
profile, $q=3/2$, and illustrates the fact that 
$\mathcal{E}_{M}(k_n) > \mathcal{E}_{K}(k_n)$ for 
$0 < k_n < k_{\rm BH}$ and $0 < q < 2$.

Using equation (\ref{eq:eigenvector_ratios}) and the expression for
the dispersion relation (\ref{eq:dispersion_relation}), it is easy to
show that the following inequalities hold for {\it each} wavenumber $k_n$
\begin{eqnarray}
\label{eq:stresses_to_energies_per_k}
\frac{\mathcal{R}_{r\phi}(k_n)}{\mathcal{E}_K(k_n)} =
\frac{-\mathcal{M}_{r\phi}(k_n)}{\,\,\mathcal{E}_M(k_n)} =
\frac{2\gamma(k_n)}{q} \le 1\,,
\end{eqnarray}
as long as $q>0$.
It immediately follows that the same inequalities
are also satisfied by the corresponding means, i.e., 
\begin{eqnarray}
 \bar{R}_{r\phi}(t) &\le& \bar{E}_K(t) \,,\\
-\bar{M}_{r\phi}(t) &\le& \bar{E}_M(t) \,.
\end{eqnarray}
This result, in turn, implies that the total mean energy associated
with the fluctuations, $\bar{E}(t) = \bar{E}_K(t) + \bar{E}_M(t)$,
sets an upper bound on the total mean stress, i.e.,
\begin{eqnarray}
\bar{T}_{r\phi}(t) \le \bar{E}(t) \,.
\end{eqnarray}
At late times during the exponential growth of the instability, the
growth of the fluctuations is dominated by the mode with
$k_n=k_{\max}$ and the mean stress $\bar{T}_{r\phi}(t)$ will tend to
the total mean energy $\bar{E}(t)$, i.e.,
\begin{eqnarray}
\lim_{t \gg 1} \bar{T}_{r\phi}(t) = \lim_{t \gg 1}\bar{E}(t) \,.
\end{eqnarray}

Furthermore, according to the first equality in equation
(\ref{eq:stresses_to_energies_per_k}), we can conclude that the ratio
of the mean magnetic to the mean kinetic energies has the same
functional dependence on the shear parameter, $q$, as does the
(negative of the) ratio between the mean Maxwell and the mean Reynolds
stresses given by equation (\ref{eq:analytical_ratio_late}), i.e.,
\begin{equation}
\lim_{t \gg 1} \frac{\bar{E}_M(t)}{\bar{E}_K(t)} = \frac{4-q}{q} \,.
\end{equation}
Therefore, the mean energy associated with magnetic fluctuations is
always larger than the mean energy corresponding to kinetic
fluctuations as long as the flow is Rayleigh-stable.

\label{lastpage}

\end{document}